\documentclass[aps,pra,twocolumn,groupedaddress]{revtex4-1}
\usepackage{amssymb,amsmath,bm,graphicx,here}

\begin{document}
\title{Gapless surface states originated from accidentally degenerate quadratic band touching in a three-dimensional tetragonal photonic crystal}
\author{Tetsuyuki Ochiai}
\affiliation{Research Center for Functional Materials, National Institute for Materials Science (NIMS), Tsukuba 305-0044, Japan}

\begin{abstract}
A tetragonal photonic crystal composed of high-index pillars can exhibit a frequency-isolated accidental degeneracy at a high-symmetry point in the first Brillouin zone. A photonic band gap can be formed there by introducing a geometrical anisotropy in the pillars.  In this gap, gapless surface/domain-wall  states emerge under a certain condition.  
We analyze their physical property in terms of an effective hamiltonian,  and a good agreement between the effective theory and numerical calculation is obtained.  
\end{abstract}

\maketitle  

\section{INTRODUCTION}

Gapless boundary states are often attributed to a certain property in the bulk states. One representative example is a quantum Hall system, where the so-called bulk-edge correspondence is established \cite{PhysRevB.43.11025,hatsugai1993cna}.
Namely, if the system is topologically nontrivial in the bulk, then the system sustains gapless boundary states. Such boundary states are generally robust against various perturbations. 
This bulk-edge correspondence rule  is found to hold in a wide variety of classical and quantum systems, including photonic \cite{haldane2008prd}, mechanical \cite{huber2016topological}, acoustic \cite{PhysRevLett.114.114301}, and magnonic \cite{PhysRevB.87.174427} systems.

Closely related to this rule, two-dimensional (2D) massive Dirac hamiltonian has the gapless domain-wall states in the system with two domains having opposite sign of the mass term \cite{PhysRevD.13.3398}.  These domain-wall states have the linear dispersion with the definite slope, irrespective of details in the domain-wall profile. Solely the relative sign of the mass term determines the slope. In this sense, the domain-wall states are topological and robust against the modulation of the domain-wall profile.   
Similar domain-wall states also exist in a 2D quadratic hamiltonian with a time-reversal violating perturbation \cite{JPSJ.82.124005}.

In optics, such boundary and domain-wall states in 2D systems can be used as a novel channel waveguide that is robust against various perturbations.  If the system breaks the time-reversal symmetry by the magneto-optical effect, resulting boundary states can form a nonreciprocal (one-way) waveguide \cite{haldane2008prd,wang2009observation,ao2009one,ochiai2009photonic}. If the time-reversal symmetry holds, they can form a helical, or in other words, polarization/spin dependent one-way waveguide \cite{hafezi2011robust,khanikaev2013photonic,PhysRevLett.110.203904,PhysRevLett.114.223901}. These waveguides are quite extraordinary and not available by conventional designs of the optical nonreciprocity \cite{pozar2009microwave,potton2004reciprocity,1468-6996-16-1-014401}.

In three-dimensional (3D) optical systems, similar boundary/domain-wall states 
are also important. There are several reasons for this statement:  1) In three spatial dimensions, a wide variety of nontrivial topology \cite{schnyder2008classification,lu2013weyl,lu2016symmetry} and thus a wide variety of optical functionality are available. 2) Optical 3D systems have the built-in photospin-orbit interaction \cite{PhysRevLett.97.193903} that is essential for possible spin-dependent light transport. 3) Ultimate photonic integrated circuits are photonic 3D architectures \cite{joannopoulos2011photonic}. 
Therefore, functional 3D optical components are definitely in order.

In this paper, we propose a way to construct optical gapless surface/domain-wall states in three spatial dimensions. We focus on a 3D tetragonal photonic crystal (PhC) composed of high-k material pillars. 
At high symmetry points in the Brillouin zone, four bands with two mutually-different doubly-degenerate representations can be accidentally degenerate quadratically.  We first design this accidental degeneracy by tuning geometries of the pillars. Then, we introduce a symmetry-breaking perturbation for the parity  with respect to the pillar axis. This causes a gap opening. In the gap, gapless surface/domain-wall states are shown to emerge.

Similar domain-wall states was reported in 3D hexagonal PhCs \cite{slobozhanyuk2016three} as 3D photonic topological insulators. There, a linear band touching of the Dirac type (having four eigenstates with two mutually degenerate bands) is argued. We extend this argument to the case of a quadratic band touching.     
A related structure of tetragonal photonic crystals was also proposed to emulate the topological crystalline insulator \cite{PhysRevB.84.195126}. However, its design strategy is very different from ours.

This paper is organized as follows. We first demonstrate numerically the emergence of the gapless surface/domain-wall states in Sec. II. 
We then argue these features by employing an effective hamiltonian description with an aid of group theory in Sec. III.  
Summary  and discussions are given in Sec. IV.

\section{NUMERICAL RESULTS}

\begin{figure}
\begin{center}
\includegraphics[width=0.45\textwidth]{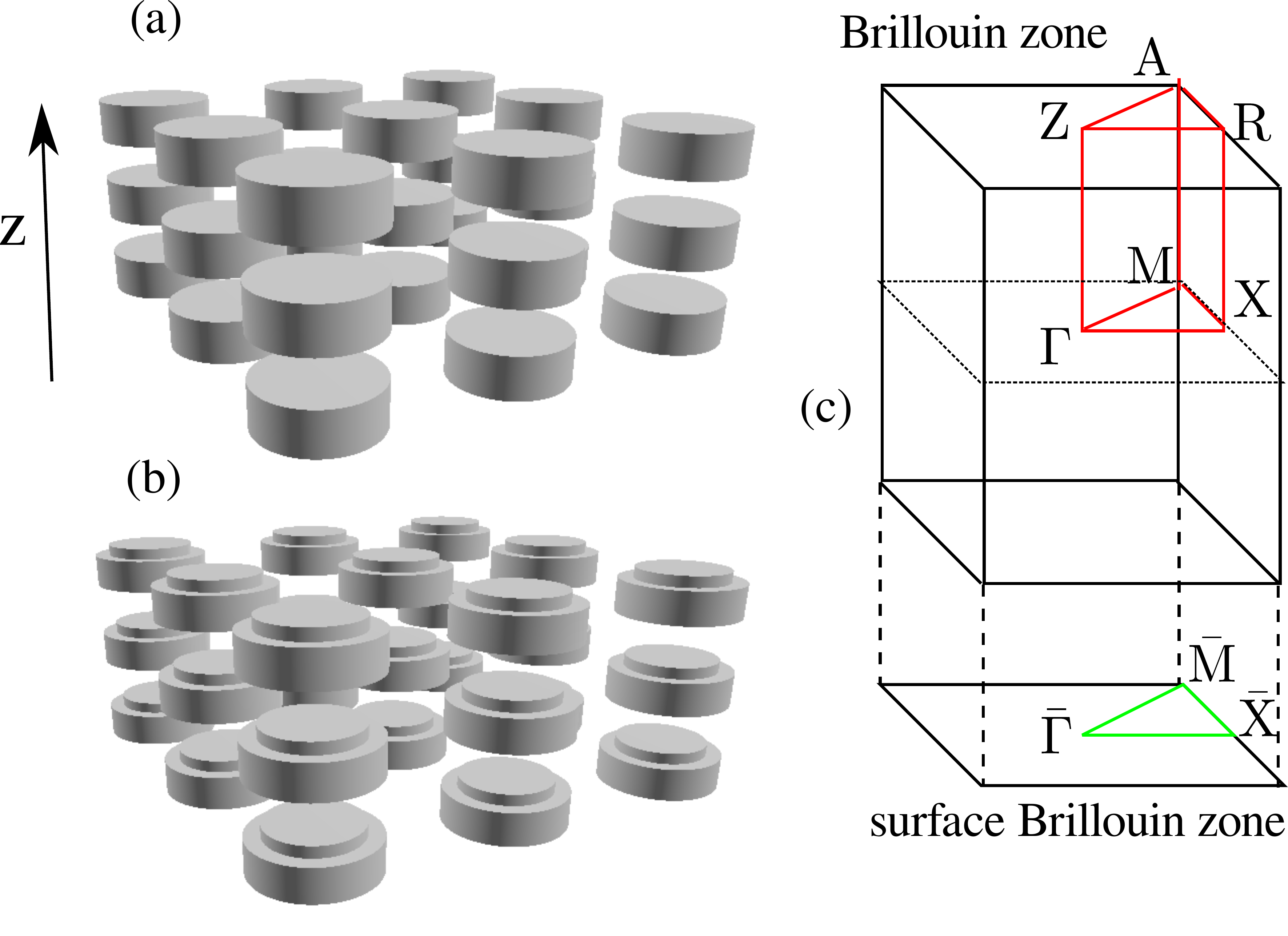}
\end{center}
\caption{\label{Fig_geometry} (a) Tetragonal photonic crystal composed of circular pillars. The pillar axis is taken to be parallel to the $z$ direction.  (b)  Tetragonal photonic crystal composed of two-step circular pillars. The parity symmetry in the $z$ direction is broken.  
(c) The first Brillouin zone for the bulk photonic crystals of (a) and (b), and the surface Brillouin zone relevant to the system without the translational invariance along the $z$ direction. Points of high symmetry are also indicated.  
} 
\end{figure}

We consider a 3D tetragonal PhC composed of high-k material pillars. A schematic illustration of the system under study is shown in Fig. \ref{Fig_geometry}. 
We first deal with simple circular pillars. Then, we consider two-step pillars that cause the parity symmetry breaking in the $z$ axis (taken as the pillar axis).
The point group of this PhC before the symmetry breaking is $D_{4h}$, which has two doubly degenerate irreducible representations, $E_g$ and $E_u$ at high symmetry points $\Gamma$, M, Z, and A in the first Brillouin zone \cite{inui1996group}.  
We can design an accidental degeneracy there between the two modes, $E_g$ and $E_u$.

\begin{figure}
\begin{center}
\includegraphics[width=0.45\textwidth]{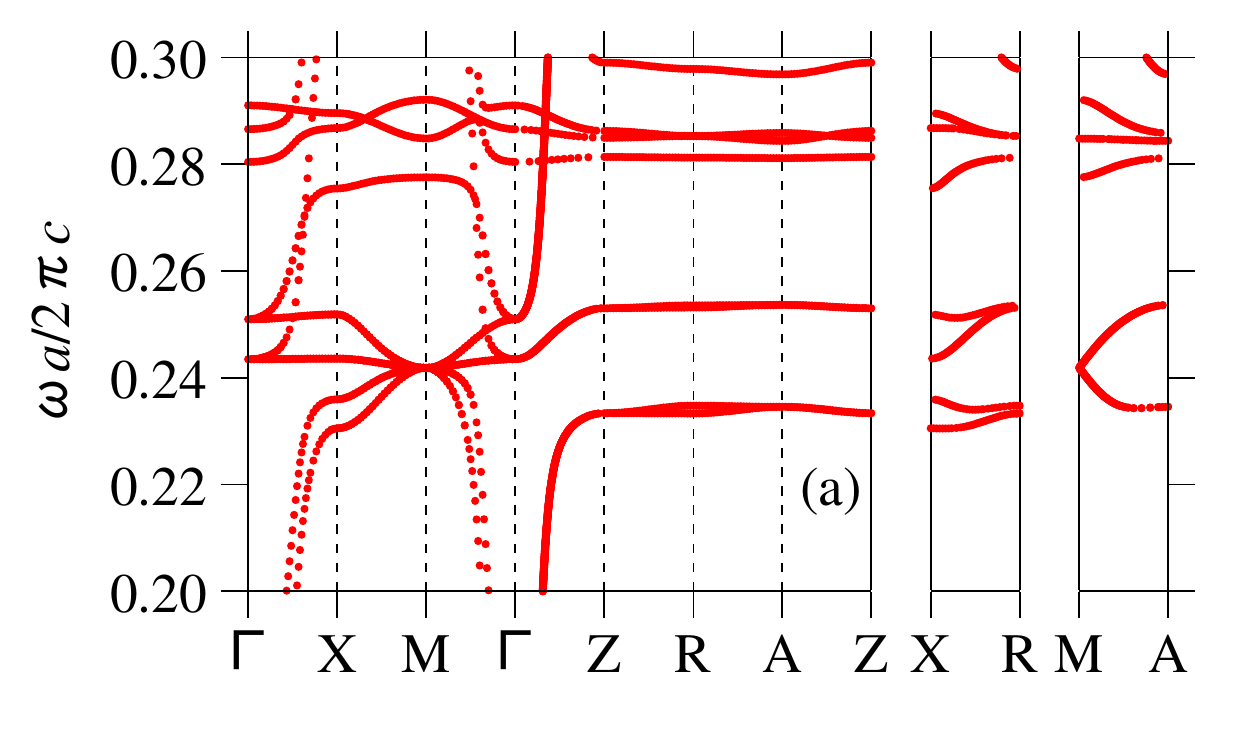}\\
\includegraphics[width=0.45\textwidth]{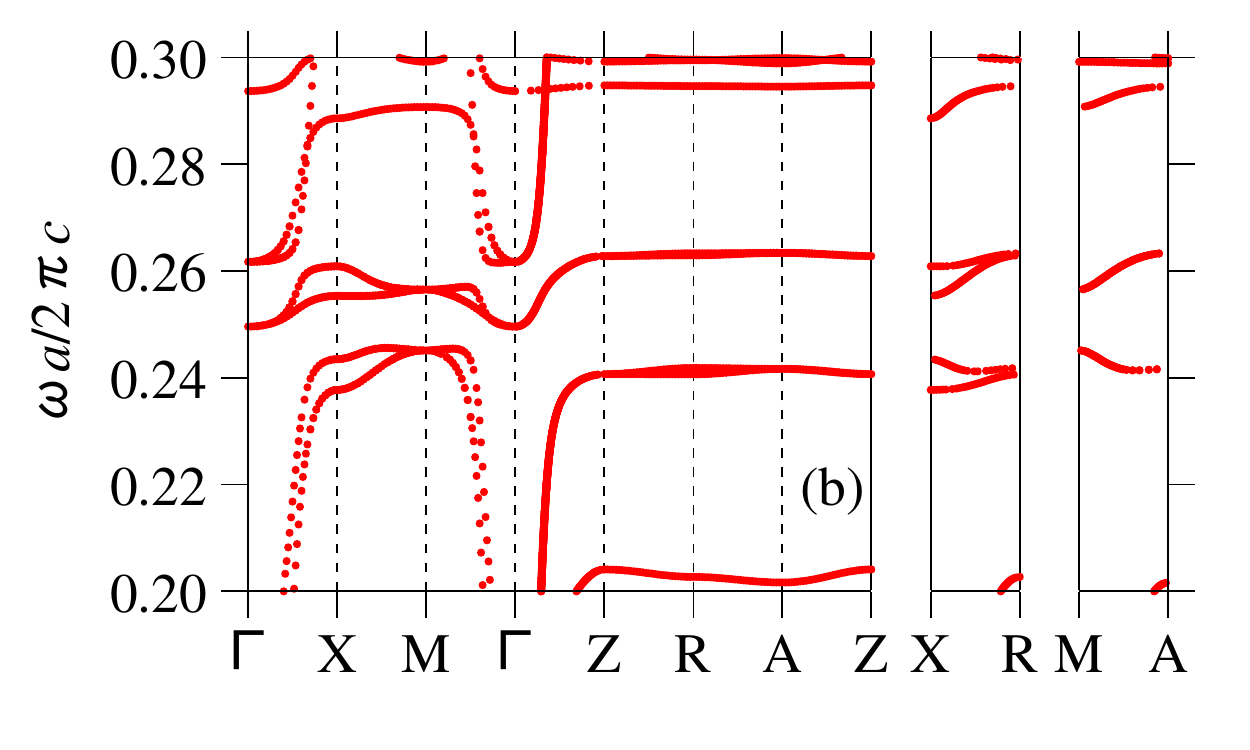}
\end{center}
\caption{\label{Fig_band} Photonic band structure of a tetragonal photonic crystal composed of  circular pillars of high-k material before (a) and after (b) introducing the $z$-parity-symmetry breaking.  In (a), the dielectric constant $\epsilon$, radius $r$, and height $h$ of the pillars are taken to be 100, $0.295a$, and $0.259a$, respectively, where $a$ is the lattice constant in plane. The lattice constant $a_z$ in the  $z$ direction is $0.5a$. In (b) the pillar is an coaxial two-step one with inner-core radius of $0.75r$ and height $h$. The outer-shell has radius $r$ and height $0.75h$. } 
\end{figure}

Figure \ref{Fig_band} (a) shows the photonic band structure of a tetragonal PhC with the accidentally degeneracy around $\omega a/2\pi c=0.24$ at the M point. 
Here, $\omega$ is the angular frequency of light, $a$ is the lattice constant of the PhC in plane, and $c$ is the light velocity in vacuum.  
The rigorous coupled-wave analysis (RCWA) is employed in the numerical calculation \cite{Noponen1994,Li1997,Tikhodeev:Y:M:G:I::66:p045102:2002,PhysRevE.67.046607}.  
The accidental degeneracy results in a quadratic band touching among the four bands in the $(k_x,k_y)$ plane. As for the $k_z$ direction (along the MA axis), the band dispersion is linear and doubly degenerate. Such a band anisotropy comes from the structural anisotropy of the tetragonal PhC.

By introducing the symmetry breaking to the pillars, the point group of the system reduces to $C_{4v}$. Both the $E_g$ and $E_u$ modes reduce to the $E$ modes of $C_{4v}$. Since the modes of the same representation repel each other, the accidentally degeneracy between the  $E_g$ and $E_u$ modes is lifted after introducing the symmetry breaking, forming a band gap there.    
We can see clearly the gap opening around $\omega a/2\pi c=0.245$ in Fig. \ref{Fig_band} (b). This gap is shown to support gapless domain-wall states.

\begin{figure}
\centerline{
\includegraphics[width=0.2\textwidth]{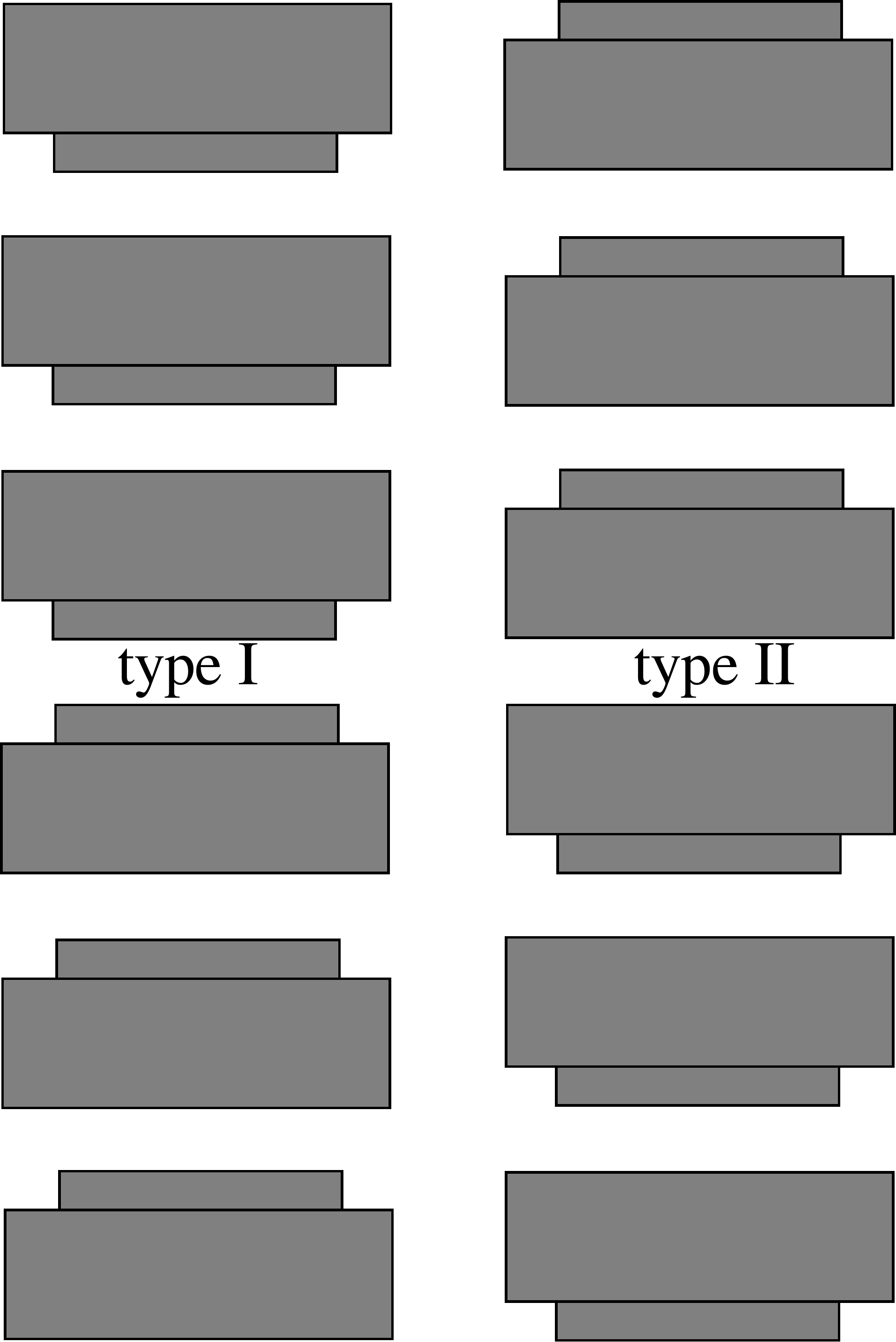}
}
\caption{\label{Fig_dwschematic} Schematic illustration of domain walls used in the numerical calculation. Two types of the domain walls are considered. } 
\end{figure}

Next, we consider a domain wall formed by two domains stacked in the $z$ direction. The two domains 
 have the opposite sign of the parity-symmetry breaking (inverted the pillar axis). Two types of the domain wall is considered as shown in Fig. \ref{Fig_dwschematic}.

\begin{figure}
\centerline{
\includegraphics[width=0.45\textwidth]{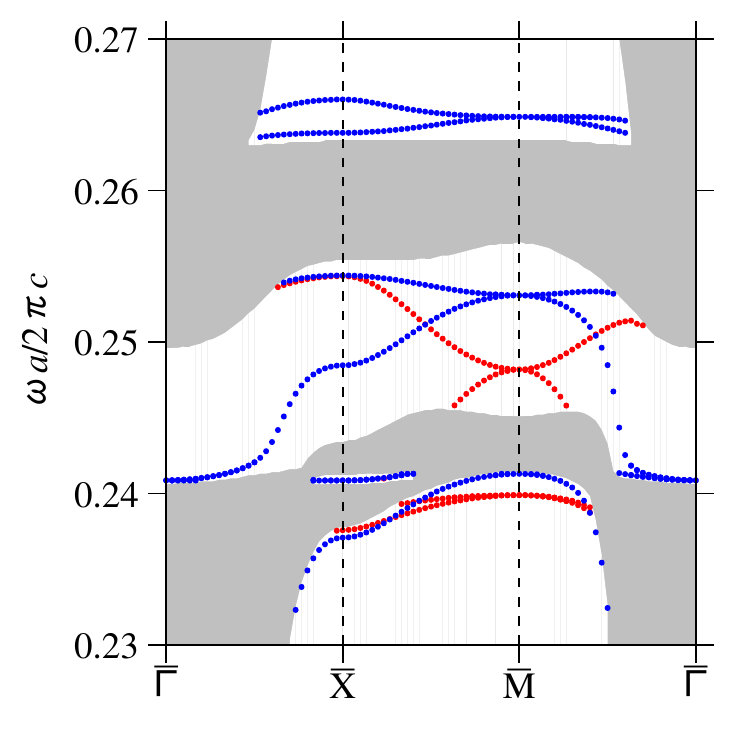}
}
\caption{\label{Fig_dwz} Dispersion relation of the domain-wall states in the composite structure with the two domains of the symmetry-broken tetragonal photonic crystal. The pillar axis of one domain is inverted from that in the other domain.  The parameters are the same as in the Fig. \ref{Fig_band} (b). The domain-wall states of type I (II) are indicated by red (blue) dot. The shaded region is the projection of the bulk band structure. } 
\end{figure}

Figure \ref{Fig_dwz} shows the band structure of the domain-wall states, evaluated by the RCWA.   
In the gap around $\omega a/2\pi c=0.245$, there are gapless domain-wall dispersion curves  forming quadratic band touching at the $\bar{\rm M}$ point in the surface Brillouin zone.    
The spatial symmetries are different between the two types. In type I, the mode symmetry at $\bar{\rm M}$ is $E_g$, whereas in type II, it is $E_u$ (note that the composite system has the $D_{4h}$ symmetry as a whole, though each domain has solely the $C_{4v}$ symmetry).

\begin{figure}
\begin{center}
\includegraphics[width=0.45\textwidth]{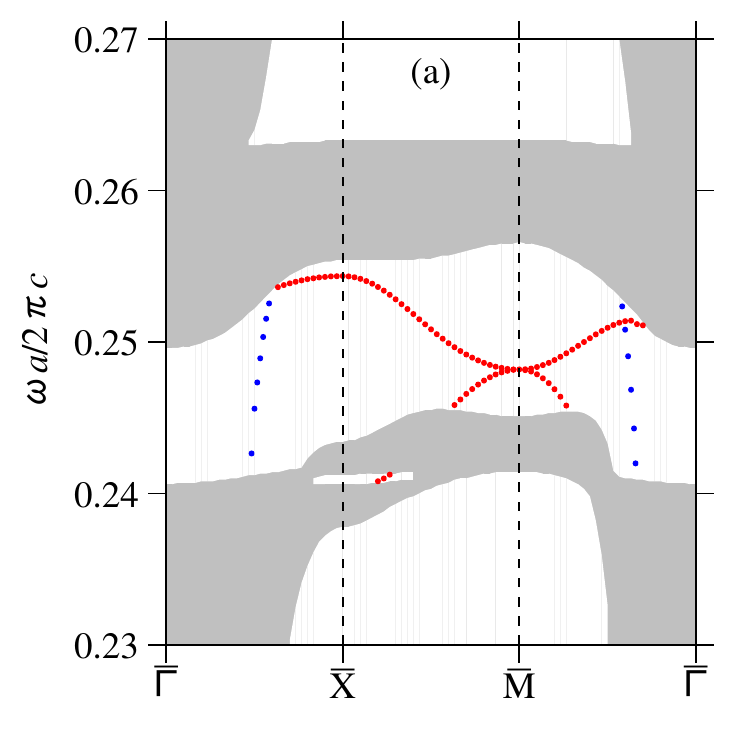}
\includegraphics[width=0.45\textwidth]{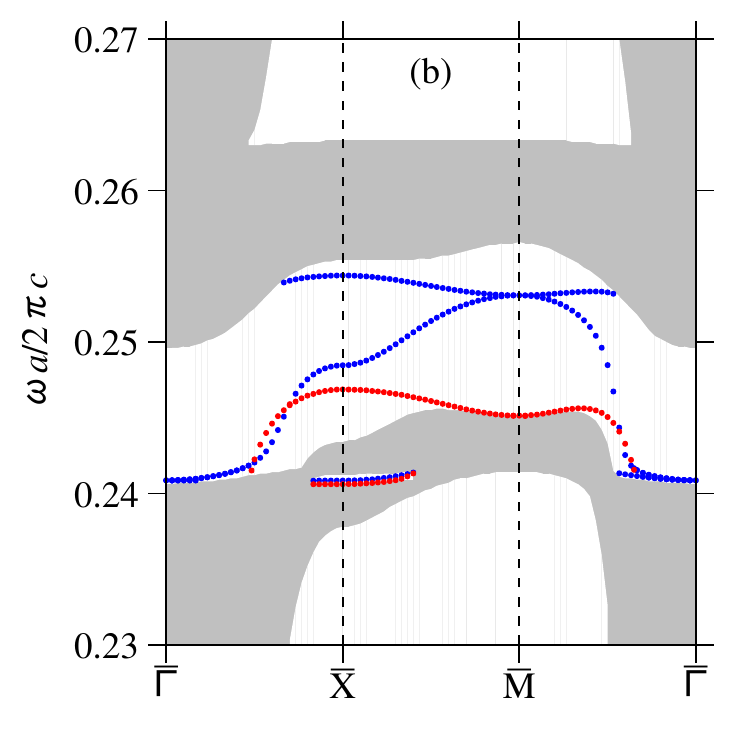}
\end{center}
\caption{\label{Fig_PECPMC} Dispersion curves of the surface states in the 32-layer-thick tetragonal photonic crystal without the $z$ parity symmetry. Projection of the bulk band structure is also plotted. The parameters are the same as in the Fig. \ref{Fig_band} (b). In (a), the PhC is capped by the perfect-electric conductor wall at the top surface. The bottom surface is open. In (b), the PhC is capped by the perfect-magnetic-conductor wall at the bottom surface. The top surface is open.The surface states at the top (bottom) are indicated by red (blue) dot.} 
\end{figure}

Similar surface states with quadratic band touching emerge in the upper surface of the single-domain finite-thick PhC of Fig. 1(b) capped by the perfect-electric-conductor (PEC) wall.  
Figure \ref{Fig_PECPMC} (a) shows such surface states in the finite-thick PhC. 
Here, the bottom surface is open, touching with outer medium of $\epsilon=1$, where $\epsilon$ is the permittivity. The surface states at the bottom are gapless.

If the perfect-magnetic-conductor (PMC) wall is placed at the bottom surface, it produces similar surface states as in the type II domain wall, as shown in Fig. \ref{Fig_PECPMC} (b). In this case, the top surface, which is open to outer medium, has the gapful dispersion of the surface states.

The quadratic band touching and gapless-ness of the domain-wall  are robust against changing the domain-wall width.  However, the surface states of the PEC/PMC wall can be gapped out by changing the distance to the wall.  If the distance is very large compared to relevant wavelength, the system can be regarded as an isolated single-domain PhC with open boundaries. In this case, the surface states at the top are fully gapped as shown in Fig. \ref{Fig_PECPMC} (b).  
In this sense, the system is not topological, and gapless-ness and robustness are limited in the domain-wall states.

\section{UNDERSTANDING VIA EFFECTIVE THEORY}

To understand the numerical results given in Sec. II,  
we consider an effective hamiltonian around the accidental degeneracy. It is generally written as \cite{landau2013quantum}
\begin{align}
&{\cal H}_{pq}=\langle \psi_0^{(0p)}|H'|\psi_0^{(0q)}\rangle \nonumber \\
&\hskip20pt +\sum_{n\ne 0}\frac{\langle \psi_0^{(0p)}|H'|\psi_0^{(n)}\rangle  
\langle \psi_0^{(n)}|H'|\psi_0^{(0q)}\rangle}{E_0^{(0)}-E_0^{(n)}}, \label{Eq_effham}\\
&H=H_0 + H',\\
&H_0|\psi^{(n)}\rangle =E_0^{(n)}|\psi^{(n)}\rangle, 
\end{align}
for a generic hamiltonian $H$ with a degeneracy in the zeroth order approximation.  
The states  $|\psi_0^{(0p)}\rangle$ ($p=1,2,..,N$) are the $N$-fold degenerate eigenstates of the zeroth order hamiltonian $H_0$ with eigenvalue $E_0^{(0)}$.  
The other eigenstates are denoted as $|\psi_0^{(n)}\rangle$ with eigenvalue $E_0^{(n)}$ ($n\ne 0$). 
The lifting of the degeneracy is described by the effective hamiltonian.

Besides, the Maxwell equation casts into a hamiltonian (diagonalization) form as  
\begin{align}
&H|\psi\rangle = E|\psi\rangle,\\
&\langle {\bm G}|H|{\bm G}'\rangle=-\eta_{{\bm G}-{\bm G}'}({\bm k}+{\bm G})\times ({\bm k}+{\bm G}')\times, \label{Eq_Maxwellham}\\
& \langle {\bm G}|\psi\rangle = {\bm h}_{{\bm G}},\quad E=\frac{\omega^2}{c^2},\\
& \frac{1}{\epsilon({\bm x})}=\sum_{\bm G}\eta_{\bm G}{\rm e}^{{\rm i}{\bm G}\cdot{\bm x}}, 
\end{align}
where $\epsilon({\bm x})$ is the periodic  permittivity in the PhC and ${\bm G}$ is a reciprocal lattice vector.  The vector ${\bm h}_{\bm  G}$ is the Fourier coefficient of the magnetic field of Bloch momentum ${\bm k}$ and angular frequency $\omega$:   
\begin{align}
& {\bm H}({\bm x},t)=\Re\left[{\bm H}({\bm x}){\rm e}^{-{\rm i}\omega t}\right],\\
& {\bm H}({\bm x})=\sum_{\bm G}{\bm h}_{\bm G}{\rm e}^{{\rm i}({\bm k}+{\bm G})\cdot {\bm x}},
\end{align}
where $\Re$ represents the real part.

To derive the effective hamiltonian, we put ${\bm k}={\bm k}_0+\delta{\bm k}$ and $\eta_{\bm G}=\eta_{\bm G}^0 + \delta \eta_{\bm G}$. Here, ${\bm k}_0$ is taken to be a high-symmetry point in the Brillouin zone, where the accidental degeneracy takes place, 
 and $\eta_{\bm G}^0$ is the Fourier coefficient of the permittivity of the PhC with the $z$ parity symmetry. The deviation $\delta \eta_{\bm G}$ is parametrized linearly by $\xi$, and satisfies $\delta \eta_{\sigma_z{\bm G}}=-\delta \eta_{\bm G}$, where $\sigma_z$ represents the $z$ parity operation.  
The zeroth-order hamiltonian is given by Eq. (\ref{Eq_Maxwellham}) at  ${\bm k}={\bm k}_0$ and $\eta_{\bm G}=\eta_{\bm G}^0$. The perturbation part is the rest of $H$, namely, $H'=H-H_0$.

Suppose that the system has the $D_{4h}$ symmetry at a high-symmetry point ${\bm k}={\bm k}_0$ in the Brillouin zone and that $E_g$ and $E_u$ modes are accidentally degenerate. The effective hamiltonian ${\cal H}$ consists of two parts, namely, the ${\bm k}\cdot{\bm p}$ part ${\cal H}_{\delta{\bm k}}$ and symmetry-breaking part ${\cal H}_\xi$. 
The  ${\bm k}\cdot{\bm p}$ part satisfies
\begin{align}
&{\cal H}_{\delta{\bm k}}=D^\dagger(A){\cal H}_{A\delta{\bm k}} D(A),\\
&D(A)={\rm Bdiag}(D_{E_g}(A),D_{E_u}(A)),
\end{align} 
where $A$ is an element of $D_{4h}$, and $D_{E_{g(u)}}(A)$ is its representation matrix for irreducible representation $E_{g(u)}$. Namely,  
\begin{align}
A{\bm E}_{E_{g(u)}}^{(p)}(A^{-1}{\bm x})=\sum_{q=1,2}{\bm E}_{E_{g(u)}}^{(q)}({\bm x})[D_{E_{g(u)}}(A)]_{qp}, 
\end{align}
where ${\bm E}_{E_{g(u)}}^{(p)}({\bm x})$ ($p=1,2$) is the electric field eigenstate of the doubly degenerate representation $E_{g(u)}$.  
The symmetry-breaking part satisfies 
\begin{align}
&{\cal H}_{\xi}=D^\dagger(A){\cal H}_{\xi} D(A) \quad (A\in C_{4v}),\\
&{\cal H}_{\xi}=-D^\dagger(\sigma_z){\cal H}_{\xi} D(\sigma_z).  
\end{align}

Using this relation, we can show that the effective hamiltonian up to the second order in $\delta {\bm k}$ and first order in $\xi$ has the following form:  
\begin{align}
&{\cal H}=
\left(\begin{array}{cc}
{\cal H}_{++} & {\cal H}_{+-}\\
{\cal H}_{-+} & {\cal H}_{--}
\end{array}\right),\label{Eq_effham2}\\
&{\cal H}_{++}= ( a_{0x} \delta{\bm k}_\|^2 + a_{0z}\delta k_z^2)\hat{1}
+a_{3x}(\delta k_x^2-\delta k_y^2) \sigma_3 \nonumber \\
& \hskip50pt + a_{1xy}\delta k_x\delta k_y \sigma_1,\\
&{\cal H}_{+-}= (a_{3z} \delta k_z + {\rm i} b_3\xi ) \sigma_3,\\
&{\cal H}_{-+}= (a_{3z} \delta k_z - {\rm i} b_3\xi ) \sigma_3,\\
&{\cal H}_{--}= ( a_{0x}' \delta{\bm k}_\|^2 + a_{0z}'\delta k_z^2)\hat{1} 
+a_{3x}'(\delta k_x^2-\delta k_y^2) \sigma_3 \nonumber \\
& \hskip50pt + a_{1xy}'\delta k_x\delta k_y \sigma_1, 
\end{align}  
where $\hat{1}$ is the $2\times 2$ unit matrix, and $\sigma_i$ ($i=1,2,3$) is the Pauli matrix. The coefficients  $a_{0x}, a_{0z}, a_{1xy}, a_{3x}, a_{3z}, a_{0x}', a_{0z}', a_{1xy}', a_{3x}'$, and $b_3$ are all real and are determined from the unperturbed eigenstates.

By diagonalizing the effective hamiltonian, four eigenvalues are obtained. 
At $\delta {\bm k}_\|=0$, we have two doubly degenerate eigenvalues of 
a massive Dirac type:  
\begin{align}
\varepsilon=\pm \sqrt{(a_{3z}\delta k_z)^2+(b_3\xi)^2}.
\end{align}
At $\delta k_z=\xi=0$, the four eigenvalues becomes
\begin{align}
&\varepsilon=a_{0x}\delta{\bm k}_\|^2 \pm \sqrt{a_{3x}^2(\delta k_x^2-\delta k_y^2)^2+a_{1xy}^2\delta k_x^2\delta k_y^2},\\
&\hskip20pt a_{0x}'\delta{\bm k}_\|^2 \pm \sqrt{a_{3x}'^2(\delta k_x^2-\delta k_y^2)^2+a_{1xy}'^2\delta k_x^2\delta k_y^2},
\end{align} 
which are sticked together at $\delta{\bm k}_\|=0$.  
These properties are fully consistent with the band diagram in Fig. \ref{Fig_band}.
We should point out that there is the linear term of $\delta k_z$ in the effective hamiltonian, whereas linear terms of $\delta {\bm k}_\|$ are absent. 
Since the linear term is dominating in the vicinity of $\delta k_z=0$, 
 the quadratic term in $\delta k_z$ is fairly neglected there.

We consider a composite structure with two domains characterized by opposite sign of parameter $\xi$.  The other parameters are common in the two domains. The effective hamiltonian in the composite structure is obtained by replacing $\delta k_z$ by $-{\rm i}\partial /\partial z$ and $\xi$ by $\xi(z)$. Typically, $\xi(z)$ is given by $\xi_0 \tanh(z/w)$, where $z=0$ is the center coordinate of the domain wall and $w$ is its width.  
The eigenvalue equation for the domain-wall states is thus given by  
\begin{align}
{\cal H}(k_z\to -{\rm i}\partial/\partial z)\psi(z)=\varepsilon \psi(z). 
\end{align}

We can show that this equation supports the domain wall states with 
\begin{align}
&\psi(z)=\left(\begin{array}{c}
\psi_+\\
0
\end{array}\right){\rm e}^{-\frac{b_3}{a_{3z}}\int^z {\rm d}z'\xi(z')},\\
&{\cal H}_{++}\psi_+=\varepsilon\psi_+,\\
&\varepsilon=a_{0x}\delta{\bm k}_\|^2 \pm \sqrt{a_{3x}^2(\delta k_x^2-\delta k_y^2)^2+a_{1xy}^2\delta k_x^2\delta k_y^2},
\end{align}
provided $(b_3/a_{3z})\xi_0 >0$. 
On the other hand, if $(b_3/a_{3z})\xi_0 <0$, we have  
\begin{align}
&\psi(z)=\left(\begin{array}{c}
0\\
\psi_-
\end{array}\right){\rm e}^{\frac{b_3}{a_{3z}}\int^z {\rm d}z'\xi(z')},\\
&{\cal H}_{--}\psi_-=\varepsilon\psi_-,\\
&\varepsilon=a_{0x}'\delta{\bm k}_\|^2 \pm \sqrt{a_{3x}'^2(\delta k_x^2-\delta k_y^2)^2+a_{1xy}'^2\delta k_x^2\delta k_y^2}.
\end{align}
Here, we neglect the $\delta k_z^2$ term.

These eigenvalues of the domain-wall states are the same as those of the bulk modes without the symmetry breaking ($\xi=0$) at $\delta k_z=0$, within the effective theory. In fact, in the numerical calculation given in Sec. II, the dispersion curves of the domain-wall states around $\omega a/2\pi c=0.245$ in Fig. \ref{Fig_dwz} are quite similar in shape to the dispersion curves of the bulk modes around $\omega a/2\pi c=0.24$ in  Fig. \ref{Fig_band} (a),  after reducing the relative splitting of the domain-wall dispersion curves at $\bar{\rm M}$. 
First of all, they both exhibit the gapless band structures. 
Second, the band curvatures around $\bar{\rm M}$ of the domain-wall states 
have the same trend as the curvature around ${\rm M}$ of the bulk modes. 
Third, the spatial symmetries have the same set ($E_g$ and $E_u$). 
The relative splitting can be explained by fully taking account of the $\delta k_z^2\to -\partial^2/\partial z^2$ term. 
In this way, a good agreement between the effective theory and numerical calculation is obtained.

We also note that the surface states of Fig. \ref{Fig_PECPMC} can be also explained within the effective hamiltonian.  There, the PEC or PMC boundary condition is imposed. In terms of the effective theory, the radiation field ${\bm F}(={\bm E},{\bm B},{\bm D},{\bm H})$ is written as the superposition of the $E_g$ and $E_u$ modes of the unperturbed hamiltonian: 
\begin{align}
{\bm F}({\bm x})=\sum_{p=1,2}[\psi_+^{(p)}(z) {\bm F}_{E_g}^{(p)}({\bm x}) + \psi_-^{(p)}(z) {\bm F}_{E_u}^{(p)}({\bm x})].  
\end{align}
If we put the PEC (PMC) wall at a mirror plane ($z=z_0$) of the unperturbed tetragonal PhC as assumed in Fig. \ref{Fig_PECPMC}, we must have $\psi_{-(+)}(z_0)=0$. This is because, by symmetry, the $E_g$ mode has $E_x=E_y=H_z=0$ at $z=0$, and the $E_u$ mode has $H_x=H_y=E_z=0$  at $z=0$. The PEC (PMC) boundary condition imposes $E_x=E_y=0$ ($H_x=H_y=0$) for the superposed radiation field at the wall, so that $\psi_{-(+)}(z_0)=0$ is derived.

If $(b_3/a_{3z})\xi >0$, the effective hamiltonian has the following eigenstates 
\begin{align}
&\psi(z)=\left(\begin{array}{c}
\psi_+\\
0
\end{array}\right){\rm e}^{-\frac{b_3}{a_{3z}}\xi z},\label{Eq_PECb}\\
&\varepsilon=a_{0x}\delta{\bm k}_\|^2 \pm \sqrt{a_{3x}^2(\delta k_x^2-\delta k_y^2)^2+a_{1xy}^2\delta k_x^2\delta k_y^2},\label{Eq_PECb2}
\end{align}
for the PEC wall at the bottom, and 
\begin{align}
&\psi(z)=\left(\begin{array}{c}
0\\
\psi_-
\end{array}\right){\rm e}^{\frac{b_3}{a_{3z}}\xi z},\label{Eq_PMCt}\\
&\varepsilon=a_{0x}'\delta{\bm k}_\|^2 \pm \sqrt{a_{3x}'^2(\delta k_x^2-\delta k_y^2)^2+a_{1xy}'^2\delta k_x^2\delta k_y^2},\label{Eq_PMCt2}
\end{align}
for the PMC wall at the top.

Otherwise, if $(b_3/a_{3z})\xi <0$, the eigenstates become Eq. (\ref{Eq_PMCt})
for the PMC wall at the bottom, and Eq. (\ref{Eq_PECb})
for the PEC wall at the top. 
Actually, we found the latter case in Sec. II.

If the PEC or PMC wall is not put on the mirror plane, the dispersion relation of the surface states change from Eqs. (\ref{Eq_PECb2}) and (\ref{Eq_PMCt2}). In such cases, a simple argument as given above is not available because of a symmetry mismatch at the wall.

\section{Summary and discussions}
In summary, we have shown that an accidentally degenerate  quadratic band touching of four bands in a tetragonal PhC composed of circular pillars can yield gapless domain-wall states after breaking the  parity symmetry along the pillar axis.  The dispersion relation of the domain-wall states is nearly equal to the quadratic dispersion of the bulk modes of the PhC at $k_z=0$ before the symmetry breaking.  Moreover, the surface modes of a similar dispersion emerge at the boundary surface capped by the PEC or PMC wall. 
We explicitly present these property by numerical calculations of the Maxwell equation and by an effective hamiltonian description.

The anisotropic band structure of the tetragonal PhC makes the surface band structure highly surface dependent. For instance, if we cut the system at a plane normal to the $x$ direction, the translational invariance in the ($y,z$) plane holds. In this case, no surface states are predicted around $(\delta k_y,\delta k_z)=(0,0)$ under a rough estimation in  the effective hamiltonian. However, it is still possible to have nontrivial surface states that cannot be described within the effective theory.  
Detailed investigation of such surface/domain-wall states are beyond the scope of the present paper.

A gap formation by the parity symmetry breaking can occur also for the accident degeneracy  between two nondegenerate modes of $D_{4h}$, such as $A_{1g}$ and $A_{2u}$, that reduce to the same irreducible representation ($A_1$ in this case) of $C_{4v}$. 
The reduced modes repel each other because of the same irreducible representation. The effective $2\times 2$ hamiltonian around the accidentally degeneracy  becomes  
\begin{align}
&{\cal H}=(a_{0x}\delta {\bm k}_\|^2 + a_{0z}\delta k_z^2)\hat{1}
+(a_{1z}\delta k_z+b_1\xi)\sigma_1 \nonumber \\
&\hskip20pt +(a_{2z}\delta k_z+b_2\xi)\sigma_2
+(a_{3x}\delta {\bm k}_\|^2 + a_{3z}\delta k_z^2)\sigma_3.  
\end{align}
by a similar symmetry argument as in Sec. III. 
Here, the coefficients  $a_{0x}, a_{0z}, a_{1z}, b_1, a_{2z}, b_2, a_{3x}$, and $a_{3z}$ are all real.

This effective hamiltonian supports the domain-wall states with 
\begin{align}
&\varepsilon=(a_{0x}+a_{3x})\delta {\bm k}_\|^2, \\
&\psi(z)=\left(\begin{array}{c}
1\\
0
\end{array}\right){\rm e}^{-{\rm i}\frac{b_1+{\rm i}b_2}{a_{1z}+{\rm i}a_{2z}}\int^z \xi(z'){\rm d}z'},
\end{align} 
if the two domains are stacked in the $z$ direction, and 
have opposite signs of $\xi$ satisfying   
$a_{1z}b_2\xi(z\to\infty) < a_{2z}b_1\xi(z\to\infty)$. 
Again, the term quadratic in $\delta k_z$ is neglected.  
If $a_{1z}b_2\xi(z\to\infty) > a_{2z}b_1\xi(z\to\infty)$, the domain-wall states become
\begin{align}
&\varepsilon=(a_{0x}-a_{3x})\delta {\bm k}_\|^2, \\
&\psi(z)=\left(\begin{array}{c}
0\\
1
\end{array}\right){\rm e}^{-{\rm i}\frac{b_1-{\rm i}b_2}{a_{1z}-{\rm i}a_{2z}}\int^z \xi(z'){\rm d}z'}. 
\end{align}

Although these domain-wall states are robust insensitive to domain-wall profile $\xi(z)$, they are gapful in contrast to the case discussed in Secs. II and III. Therefore, the gapless domain-wall states are limited for the accidentally degeneracy between the two mutually different doubly degenerate modes.

In this paper, we have picked up a photonic system as a representative example of the quadratic accidental degeneracy. However, the argument in this paper is simply based on the group theory. Therefore, the prediction given in the paper is not limited in the photonic systems, but can emerge a wide class of wave phenomena on a certain crystal structure with the $D_{4h}\to C_{4v}$ point group.

We hope this paper stimulate further investigations of the gap and surface-state formation via accidental degeneracy together with symmetry breaking.

\begin{acknowledgments}
This work was supported by JSPS KAKENHI Grant No. 17K05507. 
\end{acknowledgments}
%

\end{document}